# Gradual localization of 5*f* states in orthorhombic UTX ferromagnets – polarized neutron diffraction study of Ru substituted UCoGe

Michal Vališka[1], Jiří Pospíšil[1,2], Anne Stunault[3], Yukiharu Takeda[4], Beatrice Gillon[5], Yoshinori Haga[2], Karel Prokeš[6], Mohsen M. Abd-Elmeguid[7], Gwilherm Nénert[3], Tetsuo Okane[4], Hiroshi Yamagami[4,8], Laurent Chapon[3], Arsene Gukasov[5], Alain Cousson[5], Etsuji Yamamoto[2] and Vladimír Sechovský[1]

[1]*Charles University in Prague, Faculty of Mathematics and Physics, Department of Condensed Matter Physics, Ke Karlovu 5, 121 16 Prague 2, Czech Republic*
[2]*Advanced Science Research Center, Japan Atomic Energy Agency, Tokai, Ibaraki, 319-1195, Japan*
[3]*Institut Laue Langevin, 71 Avenue des Martyrs, CS 20156, F-38042 Grenoble Cedex 9, France*
[4]*Condensed Matter Science Division, Quantum Beam Science Directorate, Japan Atomic Energy Agency, 1-1-1 Kouto, Sayo-cho, Sayo-gun, Hyogo 679-5148, Japan*
[5]*Laboratoire Léon Brillouin, UMR12 CEA-CNRS, Bât 563 CEA Saclay, 91191 Gif sur Yvette Cedex, France*
[6]*Helmholtz-Zentrum Berlin für Materialien und Energie, Hahn-Meitner Platz 1, D-14109 Berlin, Germany*
[7]*Universität zu Köln, II. Physikalisches Institut, Zülpicher Str. 77, 50937 Köln, Germany*
[8]*Department of Physics, Kyoto Sangyo University, Motoyama, Kamigamo, Kita-Ku, Kyoto 603-8555, Japan*

We report on a microscopic study of the evolution of ferromagnetism in the Ru substituted ferromagnetic superconductor (FM SC) UCoGe crystallizing in the orthorhombic TiNiSi-type structure. For that purpose, two single crystals with composition $UCo_{0.97}Ru_{0.03}Ge$ and $UCo_{0.88}Ru_{0.12}Ge$ have been prepared and characterized by magnetization, AC susceptibility, specific heat and electrical resistivity measurements. Both compounds have been found to order ferromagnetically below $T_C$ = 6.5 K and 7.5 K, respectively, which is considerably higher than the $T_C$ = 3 K of the parent compound UCoGe. The higher values of $T_C$ are accompanied by enhanced values of the spontaneous moment $\mu_{spont.}$ = 0.11 $\mu_B$/f.u. and $\mu_{spont.}$ = 0.21 $\mu_B$/f.u., respectively in comparison to the tiny spontaneous moment of UCoGe (about 0.07$\mu_B$/f.u.). No sign of superconductivity was detected in either compound. The magnetic moments of the samples were investigated on the microscopic scale using polarized neutron diffraction (PND) and for $UCo_{0.88}Ru_{0.12}Ge$ also by soft X-ray magnetic circular dichroism (XMCD). The analysis of the PND results indicates that the observed enhancement of ferromagnetism is mainly due to the growth of the orbital part of the uranium 5*f* moment $\mu_L^U$, reflecting a gradual localization of the 5*f* electrons with Ru substitution. In addition, the parallel orientation of the U and Co moments has been established in both substituted compounds. The results are discussed and compared with related isostructural ferromagnetic UTX compounds (T - transition metals, X – Si, Ge) in the context of a varying degree of the 5*f*-ligand hybridization.

# I. Introduction

Uranium magnetism represents a unique part of condensed matter physics and the recent discovery of uranium ferromagnetic superconductors (FM SC) keeps these materials at the forefront of scientific interest. The group of the uranium FM SC contains to date three members - $UGe_2$[1], $URhGe$[2] and $UCoGe$[3]. In particular, URhGe and UCoGe have attracted much attention in recent years because they exhibit coexistence of weak long-range ferromagnetic order and superconductivity at ambient pressure. The investigation of the magnetic field and pressure phase diagrams indicates that these compounds are close to a magnetic instability, suggesting that superconductivity in this class of materials is mediated by critical spin fluctuations associated with a ferromagnetic quantum critical point[3].

Generally, the magnetism of uranium intermetallic compounds is determined by a delicate balance between the direct 5f-5f interaction described by the Hill limit[4] and overlap of the 5f orbitals with the s, p and d states of the surrounding ions (5f-ligand hybridization)[5]. Thus, such a complex and delicate balance leads to unusual magnetic behavior and novel ground states. This is evident in the simultaneously isostructural and isoelectronic compounds UCoGe and URhGe: they do not belong to the same Ising type universality class for FM transitions although both are strongly uniaxial ferromagnets[6]. Moreover, they are characterized by a distinct difference in their response to hydrostatic pressure as increasing pressure destabilizes[7,8] and stabilizes[9,10] the FM state of UCoGe and URhGe, respectively.

Another notable magnetic feature is the observed stabilization of the originally very weak itinerant FM by substitution of the Co(Rh) site by other transition metals even when the opposite side parent compound is a paramagnet[11-13]. As a result, unusual FM domes exist in the $UCo_{1-x}T_xGe$[13,14] (T = Ru, Fe) and $URh_{1-x}T_xGe$[15,16] (T = Ru, Co) T-x phase diagrams. The observed FM domes are characterized by an increase of the Curie temperature $T_C$ or spontaneous magnetic moment $\mu_{spont.}$ with increasing initial substitutions. The development of ferromagnetism in substituted $UCo_{1-x}T_xGe$[13,14] and $URh_{1-x}T_xGe$[15,16] has a serious impact on the superconducting state, as it is immediately suppressed already by small percentages of substituent element[11]. This implies drastic changes in the ferromagnetic state and the magnitude of the spin fluctuations in the substituted systems. The complexity of the behavior is underlined by recent studies of the $UCo_{1-x}T_xGe$ and $URh_{1-x}T_xGe$ systems substituted with Fe or Ru, showing the vanishing of the FM domes around 20-40 % of substituent accompanied by development of a non-Fermi liquid (NFL) behavior[13,15,17]. However, so far there is no detailed microscopic study to explore the origin of such drastic changes both in ferromagnetism and superconductivity in the $UCo_{1-x}T_xGe$ and $URh_{1-x}T_xGe$ systems.

In this respect, we note that the FM state of parent UCoGe is rather complex as Co exhibits a moment and thus significantly contributes to the total ordered magnetic moment. Therefore, UCoGe has been the subject of recent experimental[18-22] and theoretical[23,24] efforts: in particular polarized neutron diffraction (PND) experiments on UCoGe at 0.1 K and magnetic field of 12 T showed an induced moment on the Co site that compares to the uranium moment and is antiparallel to it[21]. Macroscopic measurements were also consistent with a transition to a ferrimagnetic states in high magnetic fields[19]. However the reported polarization of the Co magnetic moment antiparallel to the U magnetic moment contrast with the behavior found in related ferromagnetic UTX (X = Al, Ga, Si, Ge) compounds, for which the U and T moments are always found to be parallel, e.g. URhSi[25], UCoAl[26,27], URhAl[28], URuAl[29]. The behavior of UCoGe is actually still controversial, since more recent studies using synchrotron radiation (X-ray magnetic circular dichroism[30,31] and Compton scattering[24]) provide strong evidence that the magnetic state of UCoGe is not anomalous and a common ferromagnetic state was found, with the U and Co magnetic moments parallel to each other.

In the present work, we study the evolution of the FM state of Ru substituted UCoGe using PND to investigate the origin of the observed dome-like shape of the *T-x* phase diagram. In particular, we have prepared two substituted UCoGe single crystals with compositions UCo$_{0.97}$Ru$_{0.03}$Ge and UCo$_{0.88}$Ru$_{0.12}$Ge, and characterized them by magnetization, AC susceptibility, specific heat and electrical resistivity measurements. We have studied them by PND to gain insight into the microscopic mechanism that underlies the anomalous increase of the magnetic moment and $T_C$. We have also performed XMCD measurements on our UCo$_{0.88}$Ru$_{0.12}$Ge single crystal to confirm the orientation of U and Co moments determined by PND. The ultimate aim of this study is to generalize the results to related isostructural ferromagnetic UTX compounds (T- transition metals, X – Si, Ge) with similar dome-like behavior.

## II. Experimental details

Single crystals were prepared using initial stoichiometric amounts of the pure elements (U-Solid State Electrotransport treated[32], Co 3N5, Ge 6N, Ru 3N5). Rather large UCo$_{0.97}$Ru$_{0.03}$Ge and UCo$_{0.88}$Ru$_{0.12}$Ge single crystals (diameter 5 mm, length 3-5cm) were grown for PND by floating zone method in an optical mirror furnace (model: FZ-T-4000-VI-VPM-PC made by Crystal System Corp. For the XMCD measurements, a somewhat smaller UCo$_{0.88}$Ru$_{0.12}$Ge crystal (diameter 2-3 mm, length 5 cm) was grown by Czochralski method in a tetra-arc furnace (Techno Search Corp.), using 99.9% purity U. All the as grown single crystals were wrapped in a Ta foil (purity 4N), sealed in a quartz tube under $10^{-6}$ mbar vacuum and annealed at 1100 °C for 1 day, then slowly cooled down to 880 °C where they remained for 14 days and finally cooled down to room temperature[33]. The quality and composition of the single crystals were checked by EDX analysis (electron microscope (SEM) Tescan Mira I LMH equipped with energy dispersive X-ray detector Bruker AXS) and Laue method. A small part of each single crystal was grinded and the crystal structure was determined at room temperature by X-ray powder diffraction using a Bruker D8 Advance diffractometer. The powder X-ray diffraction patterns were evaluated by Rietveld analysis[34] using the FullProf[35]/WinPlotr[36] programs. Several properly shaped samples were extracted from the ingots for the individual measurements using fine wire saw (South Bay Technology 810) to prevent additional stresses. Bar-shaped samples (1 mm x 1 mm x 4 mm) were used for the low temperature resistivity measurements performed in the PPMS14T (Quantum Design) with $^3$He insert down to the 350 mK. Heat capacity measurements were performed on thin plates (2 mm x 2 mm x 0.2 mm) using relaxation method in the same PPMS9T and PPMS14T devices using $^3$He insert, as well. The magnetization was measured on roughly cubic samples (2mm x 2mm x 2mm) in a MPMS7T device (Quantum Design). The same cubic samples were also used for the neutron diffraction experiments. First, we performed an experiment with unpolarized neutrons on the UCo$_{0.88}$Ru$_{0.12}$Ge single crystal at the four circle hot neutron diffractometer D9 (ILL, Grenoble) and on the UCo$_{0.97}$Ru$_{0.03}$Ge single crystal at the 5C2 diffractometer (LLB, Saclay), then with polarized neutrons on the same samples at the D3 instrument (ILL, Grenoble)[37] and 5C1 diffractometer (LLB, Saclay), respectively. For the XMCD experiments, a clean surface of the UCo$_{0.88}$Ru$_{0.12}$Ge sample was obtained by fracturing in ultra-high vacuum. The experiments were carried out at the U $N_{4,5}$ and Co $L_{2,3}$ edges at the SPring-8 BL23SU beamline[38].

## III.    Experimental results

### A. Magnetization, electrical resistivity and specific heat

The magnetization measurements of the two substituted samples (Fig. 1) reveal that the *c*-axis is the easy magnetization direction while the *a* and *b* axes are the hard ones similar to Ising-like UCoGe. Although the *a* and *b* axes are considered as the magnetically hard ones both in UCoGe and URhGe, the *ab* plane is not magnetically isotropic and the larger moment is always measured along the *b* axis referred to as the intermediate axis[20]. An important feature is that the magnetic behavior of the *a* and *b* axes are almost identical in the case of the highly substituted $UCo_{0.88}Ru_{0.12}Ge$ sample.

The maximum in the AC susceptibility data and the first derivative of the low field thermomagnetic curves measured in magnetic fields applied along the c-axis point to $T_C = 6$ K and 8.5 K for $UCo_{0.97}Ru_{0.03}Ge$ and $UCo_{0.88}Ru_{0.12}Ge$, respectively, which is in good agreement with the previous study of polycrystalline samples[39]. We have also estimated the values of $T_C$ from the analysis of the Arrott plots[40] and obtained the same results.

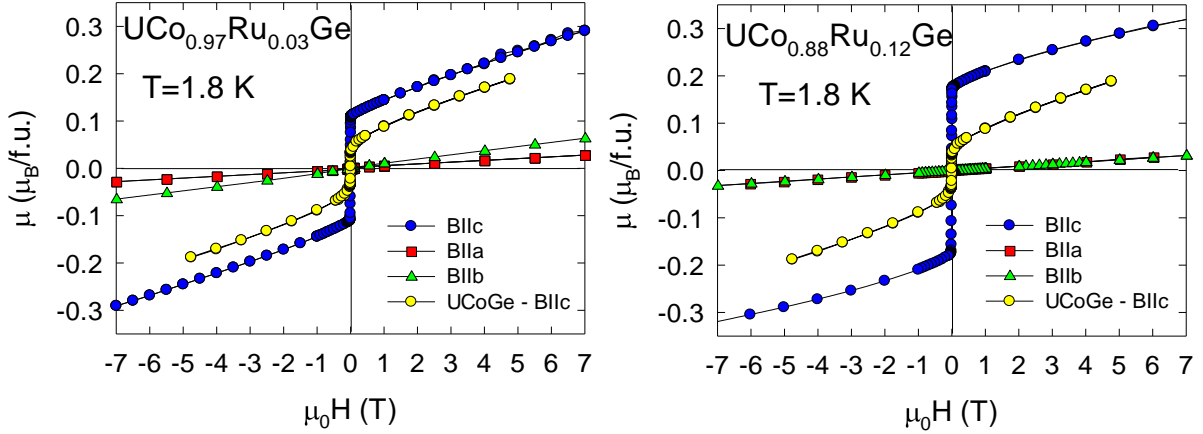

FIG. 1. (Color online) Magnetization loops for the $UCo_{0.97}Ru_{0.03}Ge$ (left) and $UCo_{0.88}Ru_{0.12}Ge$ (right) single crystals. The magnetization loop of the parent UCoGe is taken from Ref.[32] is also included for comparison. Magnetization data along the *a* and *b* axis overlap in the $UCo_{0.88}Ru_{0.12}Ge$.

The shape of the magnetization loops along the easy axis also shows a clear evolution with Ru substitution from the soft ferromagnetic behavior of UCoGe to the more rectangular behavior (gradually lower high-field susceptibility) which has been observed in URhGe[41] or URhSi[42]. The magnetic moment of UCoGe grows continuously with no sign of saturation even to 53 T. This was attributed to still-unquenched magnetic fluctuations and/or crystal-field effect[20]. We thus suggest that significantly frozen magnetic fluctuations of the U moments are responsible for the observed rectangular shape of the magnetization loops of the substituted compounds. The spontaneous magnetic moment $\mu_{spont.}$ considerably increases from 0.07 $\mu_B$/f.u. in UCoGe to 0.11 $\mu_B$/f.u. in $UCo_{0.97}Ru_{0.03}Ge$ and 0.21 $\mu_B$/f.u. in $UCo_{0.88}Ru_{0.12}Ge$. $UCo_{0.88}Ru_{0.12}Ge$ is close to the optimum concentration where the maximum $T_C$ and $\mu_{spont.}$ are achieved in the $UCo_{1-x}Ru_xGe$ system[39]. Consequently, the hysteresis increases from the initial value of ~4 mT for UCoGe up to ~5.8 mT in $UCo_{0.88}Ru_{0.12}Ge$.

The electrical resistivity along the *c* axis shows a pronounced maximum at about ~40 K as in UCoGe[32, 43] (Fig. 2). The maximum coincides with $T^*$ found by NMR study where longitudinal FM fluctuations develop along the c-axis[43]. The maxima in the substituted systems seem to be weaker and broader than in UCoGe. Experiments in applied magnetic fields would be necessary to check whether the maxima also vanish at high fields as they do in UCoGe. The electrical resistivity data also exhibit an anomaly corresponding to the Curie temperature with a shoulder at 6.5 and 7.7 K for $UCo_{0.97}Ru_{0.03}Ge$ and $UCo_{0.88}Ru_{0.12}Ge$, respectively.

It is also apparent from Fig. 2 that increasing the Ru concentration yields more substitutional disorder in the crystal lattice which consequently leads to a reduction of the RRR (residual resistivity ratio). This is consistent with the results reported for polycrystalline $UCo_{1-}$ $_x$Ru$_x$Ge[14] and UCo$_{1-x}$Fe$_x$Ge[13] samples. The disorder responsible for the low RRR can be of crystallographic and/or magnetic origin. We did not detect any sign of superconductivity in either sample (even in UCo$_{0.97}$Ru$_{0.03}$Ge, which has the smallest substitution) down to 0.35 K.

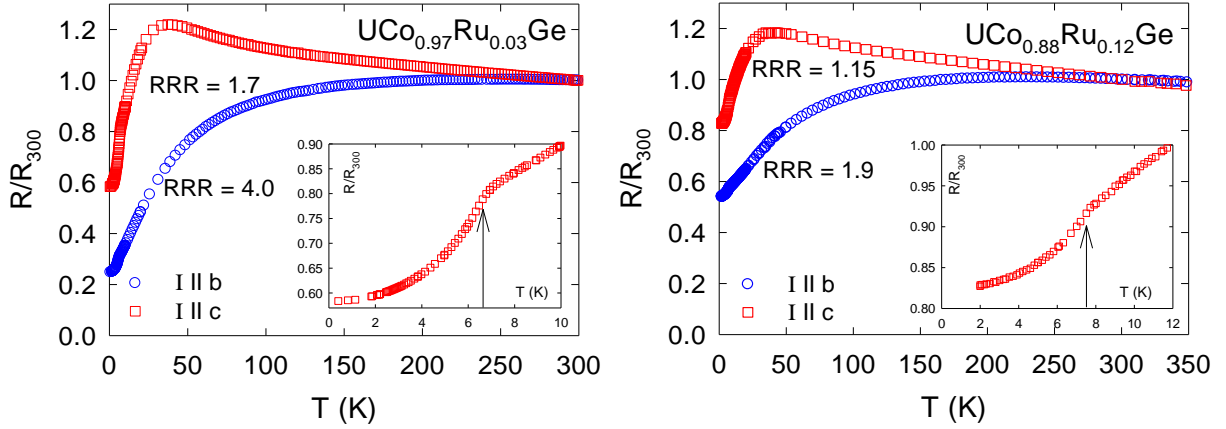

FIG. 2. (Color online) Temperature dependence of the electrical resistivity measured along the *b* and *c* axes of the UCo$_{0.97}$Ru$_{0.03}$Ge (left) and UCo$_{0.88}$Ru$_{0.12}$Ge (right) single crystals. The insets show the low temperature region, the arrows point to the anomaly at $T_C$.

The transition to the FM state is also clearly indicated by the anomaly in the specific heat (Fig. 3). The deduced $T_C$ values of 6.5 and 7.5 K for the UCo$_{0.97}$Ru$_{0.03}$Ge and UCo$_{0.88}$Ru$_{0.12}$Ge, respectively are in agreement with those obtained from magnetic and resistivity measurements. We have analyzed the $\gamma$ coefficients from the low-temperature specific heat data by linear extrapolation of the $C_p/T$ vs T$^2$ to 0 K. The obtained value of 64 mJ/molK$^2$ for UCo$_{0.97}$Ru$_{0.03}$Ge is slightly higher than that of UCoGe (57 mJ/molK$^2$)[3]. A much higher value of 79 mJ/molK$^2$ was estimated in the case of UCo$_{0.88}$Ru$_{0.12}$Ge. Similar increase of the $\gamma$ coefficients was detected in the heat capacity behavior of URh$_{1-x}$Ru$_x$Ge and explained by XPS[44] as due to an increase of the DOS at $E_F$ from enhanced 5*f*-4*d* hybridization. Stabilization of ferromagnetism with increasing Ru concentration can also be deduced from the growth of the jump at the FM transition observed in the heat capacity data (Fig. 3) and the associated increase of the magnetic entropy from $S_{mag}$ = 0.06Rln2 to $S_{mag}$ = 0.12Rln2 for UCo$_{0.97}$Ru$_{0.03}$Ge and UCo$_{0.88}$Ru$_{0.12}$Ge, respectively. The behavior of the heat capacity will be further discussed in detail in part IV and linked with the behaviors of the other related TiNiSi-type UTX compounds.

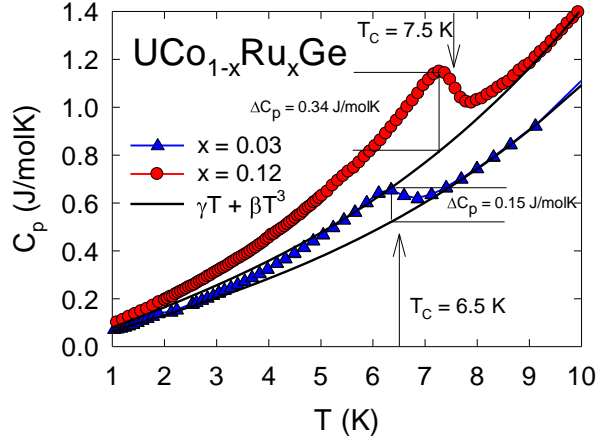

FIG. 3. (Color online) Temperature dependence of the heat capacity of the $UCo_{0.97}Ru_{0.03}Ge$ and $UCo_{0.88}Ru_{0.12}Ge$ single crystals. The arrows mark the Curie temperatures.

### B. Polarized neutron diffraction (PND)

PND is a powerful technique which can give valuable information on the spin density within the unit cell. Since the technique consists in measuring the flipping ratios, giving access to the ratios of the magnetic and nuclear structure factors, an accurate knowledge of the crystal structure is mandatory to extract reliable magnetic structure factors. In the case of $UCo_{0.88}Ru_{0.12}Ge$ the structure was determined at 11 K (i.e. above $T_C$) from the measurement of over 350 non-equivalent reflections at the D9 diffractometer of the ILL. The measured integrated intensities were analyzed using the FullProf[35]/WinPlotr[36] software using anisotropic extinction corrections. A comparison of the squares of the measured and calculated intensities is plotted in Fig. 4 showing very good agreement between the measured and calculated values.

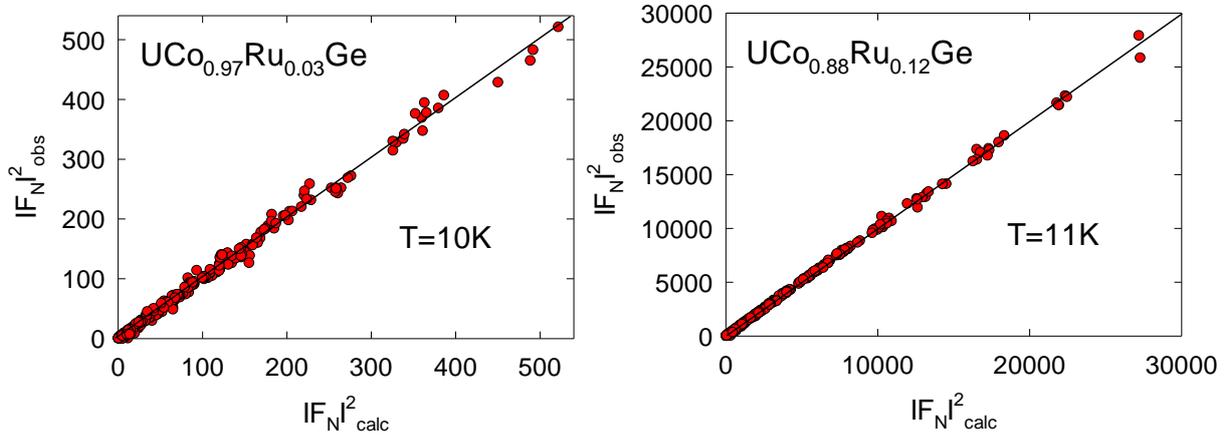

FIG. 4. (Color online) Comparison of the measured and calculated intensities. The $UCo_{0.97}Ru_{0.03}Ge$ data were taken at 10 K at the 5C-2 diffractometer of the LLB, and the $UCo_{0.88}Ru_{0.12}Ge$ data at 11 K at the D9 diffractometer of the ILL.

The data refinement confirms that both compounds crystallize in the orthorhombic TiNiSi-type structure as does the parent compound[45]. The fit also confirms that Co and Ru share only the characteristic transition metal *4c* site. The fitted Ru concentration, $x = 13.4 \pm 0.8\%$ of Ru is statistically consistent with the nominal value. The obtained cell parameters together with the fraction coordinates are listed in Tables I and II.

TABLE I. Unit cell parameters of the UCo$_{0.97}$Ru$_{0.03}$Ge UCo$_{0.88}$Ru$_{0.12}$Ge single crystals.

| Composition | Space group | a (Å) | b (Å) | c (Å) | Volume (Å$^3$) |
|---|---|---|---|---|---|
| UCo$_{0.97}$Ru$_{0.03}$Ge | Pnma | 6.8437(6) | 4.2097(4) | 7.2400(9) | 208.583 |
| UCo$_{0.88}$Ru$_{0.12}$Ge | Pnma | 6.7998(3) | 4.2104(2) | 7.2744(4) | 208.265 |

TABLE II. Evaluated fraction coordinates of the UCo$_{0.97}$Ru$_{0.03}$Ge and UCo$_{0.88}$Ru$_{0.12}$Ge single crystals from neutron diffraction data. Coefficients of the extinction parameter q(hkl) = $q_1h^2+q_2k^2+q_3l^3+q_4hk+q_5hl+q_6kl$ are listed as well.

| UCo$_{0.97}$Ru$_{0.03}$Ge | x (r.l.u.) | y (r.l.u.) | z (r.l.u.) | occ. |
|---|---|---|---|---|
| U | 0.01051(8) | 0.25000 | 0.70623(9) | 1 |
| Co | 0.28708(29) | 0.25000 | 0.41677(32) | 0.964 (8) |
| Ru | 0.28708(29) | 0.25000 | 0.41677(32) | 0.036 (8) |
| Ge | 0.19351(12) | 0.25000 | 0.08695(11) | 1 |
| Displacement temp. factors | $\beta_{11}$ | $\beta_{22}$ | $\beta_{33}$ | $\beta_{13}$ |
| U | 5.1(9) | 22(3) | 7(1) | -0.5(1) |
| Co | 24(3) | 25(9) | 11(3) | -0.4(2) |
| Ru | 24(3) | 25(9) | 11(3) | -0.4(2) |
| Ge | 10(1) | 27(3) | 11(1) | 0.3(2) |
| Extinction correction | $q_1$=0.31(4) | $q_2$=0.6(1) | $q_3$=0.12(2) | $q_{4,5,6}$=0 |

| UCo$_{0.88}$Ru$_{0.12}$Ge | x (r.l.u.) | y (r.l.u.) | z (r.l.u.) | occ. |
|---|---|---|---|---|
| U | 0.01031(7) | 0.25000 | 0.70547(8) | 1 |
| Co | 0.28375(23) | 0.25000 | 0.41618(24) | 0.866(8) |
| Ru | 0.28375(23) | 0.25000 | 0.41618(24) | 0.134(8) |
| Ge | 0.19239(8) | 0.25000 | 0.08648(9) | 1 |
| Displacement temp. factors | $\beta_{11}$ | $\beta_{22}$ | $\beta_{33}$ | $\beta_{13}$ |
| U | 5(1) | 25(3) | 10(1) | -0.7(2) |
| Co | 22(5) | 30(10) | 11(3) | 2(1) |
| Ru | 22(5) | 30(10) | 11(3) | 2(1) |
| Ge | 13(1) | 27(3) | 10(1) | 2.0(8) |
| Extinction correction | $q_1$=0.29(3) | $q_2$=0.39(4) | $q_3$=-0.27(4) | $q_{4,5,6}$=0 |

The polarized neutron diffraction experiment on UCo$_{0.88}$Ru$_{0.12}$Ge, carried out at the D3 diffractometer was performed at 1.65 K, well below the ordering temperature and in magnetic fields of 1 T and 9 T applied along the *c* axis. Flipping ratios were collected up to sin θ/λ = 0.9 Å. For UCo$_{0.97}$Ru$_{0.03}$Ge, the experiment was carried out at the 5c1 diffractometer of LLB, in a 7 T applied magnetic field along the *c* axis, and flipping ratios were collected up to sin θ/λ = 0.65 Å.

Spin densities were deduced from the obtained magnetic structure factors through a maximum entropy reconstruction[46-48]. The whole unit cell was divided into 50 x 50 x 50 =

125000 smaller cells before the computation of the distribution of the magnetic moments. The reconstructions were started from a flat magnetization distribution with a total moment in the unit cell equal to the experimental bulk magnetization measured in the same conditions. The results (i.e. magnetization density maps with the highest probability) are shown in Fig. 5.

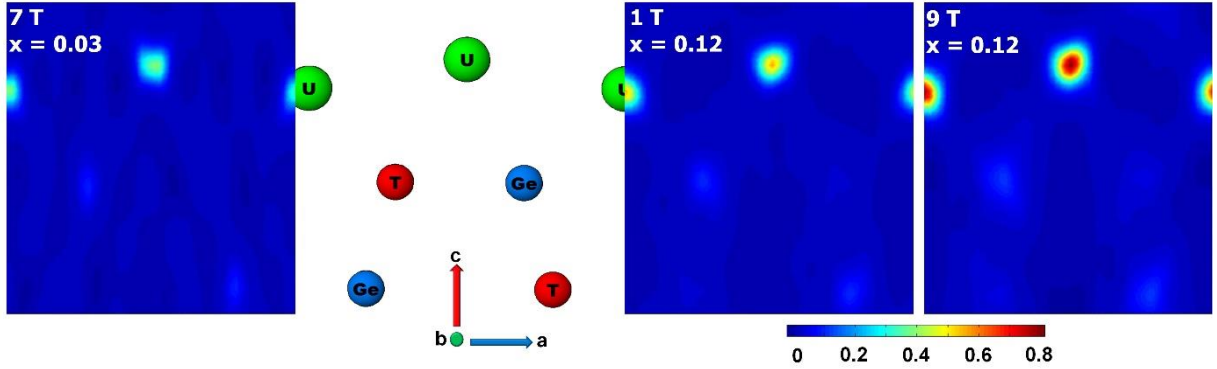

FIG. 5. (Color online) Magnetization densities in the (x, y=0.25, z) plane for $UCo_{0.88}Ru_{0.12}Ge$ in applied magnetic fields of 1 T and 9 T ∥ c and $UCo_{0.97}Ru_{0.03}Ge$ in applied magnetic field of 7 T ∥ c. The scale of all maps is m$\mu_B$/cell. This plane contains all the sites of the structure. T labels the Co/Ru site.

The Co magnetic moments are clearly oriented parallel to the U moments in both compounds, contrary to what is observed in the parent compound UCoGe[21].

We have integrated the magnetization densities in a defined volume to estimate the absolute value of the magnetic moments centered on the U and Co ions. For this purpose we choose simple spheres centered on the atomic positions. The results are summarized in Table III.

We estimated the spin and orbital components of the magnetic moment on each ion in the dipole approximation, using the FullProf[35]/WinPlotr[36] software. Our model involves the magnetic moments born at the U and Co ions. The spherical integrals for both possible ionic states of uranium ($U^{3+}$ and $U^{4+}$)[49] are similar, which disqualifies this method for the determination of the valence of the U ion. For our refinement we used a $U^{3+}$ form factor in the form $f_m(s) = W_0\langle j_0(s)\rangle + W_2\langle j_2(s)\rangle$, where $s = \sin\theta/\lambda$ is the scattering vector. The $\langle j_0(s)\rangle$ and $\langle j_2(s)\rangle$ spherical integrals are taken from Ref.[50]. The parameters $W_0$ and $W_2$ are related to the spin and orbital components of the magnetic moment: $\mu_L = W_2$, $\mu_S = W_0 - W_2$. We assumed that magnetic moments on the T=Co/Ru site was only carried by the Co ions, as confirmed by XMCD measurements (see below). For the Co ions, only the spin part of the magnetic moment was considered. The results are summarized in Table III. The fits confirm the positive magnetic density both on the U and Co ions and also the anti-parallel alignment of the orbital $\mu_L^U$ and spin $\mu_S^U$ components of the U moment.

TABLE III. Components of the magnetic moment on the U and Co(Ru) positions from the refinement of the polarized neutron diffraction data. Ratios between the orbital and spin moments on U and between the spin moments on Co and U. *Values obtained from the integration of the magnetization density maps and **values obtained from the fit of the magnetic structure factors in the dipole approximation. The table also gives the free U ion values for comparison.

| composition | $\mu_0 H$ | $\mu_L^U$ | $\mu_S^U$ | $\mu_{tot.}^U$ | $\mu_S^{Co}$ | $\mu_{tot.}$ | $|\mu_L^U/\mu_S^U|$ | $|\mu_S^{Co}/\mu_S^U|$ |
|---|---|---|---|---|---|---|---|---|
| $x = 0.03$* | 7 T | | | 0.09(1) | 0.017(3) | 0.11(1) | | |
| $x = 0.03$** | 7 T | 0.23(3) | -0.12(3) | 0.11(1) | 0.025(8) | 0.13(2) | 1.9(7) | 0.2(1) |
| $x = 0.12$* | 1 T | | | 0.11(1) | 0.07(1) | 0.18(2) | | |
| $x = 0.12$** | 1 T | 0.280(6) | -0.15(1) | 0.13(2) | 0.051(6) | 0.18(3) | 1.9(2) | 0.34(5) |
| $x = 0.12$* | 9 T | | | 0.26(3) | 0.08(1) | 0.34(4) | | |
| $x = 0.12$** | 9 T | 0.457(7) | -0.20(1) | 0.25(2) | 0.069(7) | 0.32(3) | 2.3(1) | 0.34(5) |
| $U^{3+}$ free ion[25] | | 5.585 | -2.169 | 3.416 | | | 2.6 | |
| $U^{4+}$ free ion[25] | | 4.716 | -1.432 | 3.284 | | | 3.3 | |

We have also analyzed UCo$_{0.88}$Ru$_{0.12}$Ge by XMCD in the ferromagnetic state at $T$ = 5.5 K and $H$ = 10 T. The quantitative analysis using the sum rules[50, 51] is very delicate due to the overlap of the U- $N_4$ and the Co $L_3$ edges[51]. This analysis confirms the PND results that the U and Co moments point in the same direction, as they do in UCoAl[27]. At the Ru $M_{2,3}$ edge, no XMCD signal was detected within the experimental accuracy, which denotes that the contribution from the Ru ion to the total magnetism is very weak at most.

## IV. Discussion

In the following we discuss the general trends responsible for the behavior of TiNiSi-type UTX ferromagnets and focus on the anomalous development of ferromagnetic domes in the *T-x* phase diagrams of the substituted systems. This is based on the analysis of the PND data which allowed us to resolve the macroscopic magnetization into contributions from each ion and thereby analyze the microscopic origin of the ferromagnetism.

The magnetization distribution was obtained by maximum entropy reconstruction, showing that the majority of the bulk magnetic moment is located on the uranium site, while a weaker parallel moment is detected on the transition metal site. The existence of magnetic moments on the U and Co sites with parallel orientations was confirmed on UCo$_{0.88}$Ru$_{0.12}$Ge by a soft X-ray XMCD measurement.

In the dipole approximation, the magnetic form factor on the uranium site was decomposed into orbital $\mu_L^U$ and spin $\mu_S^U$ contributions. The orbital part $\mu_L^U$ is the leading part in both compounds and is parallel to the applied magnetic field. The weaker spin part, $\mu_S^U$, is coupled antiparallel to $\mu_L^U$, similar to what has been observed in many uranium-based compounds[25, 26] including UCoGe[21, 24].

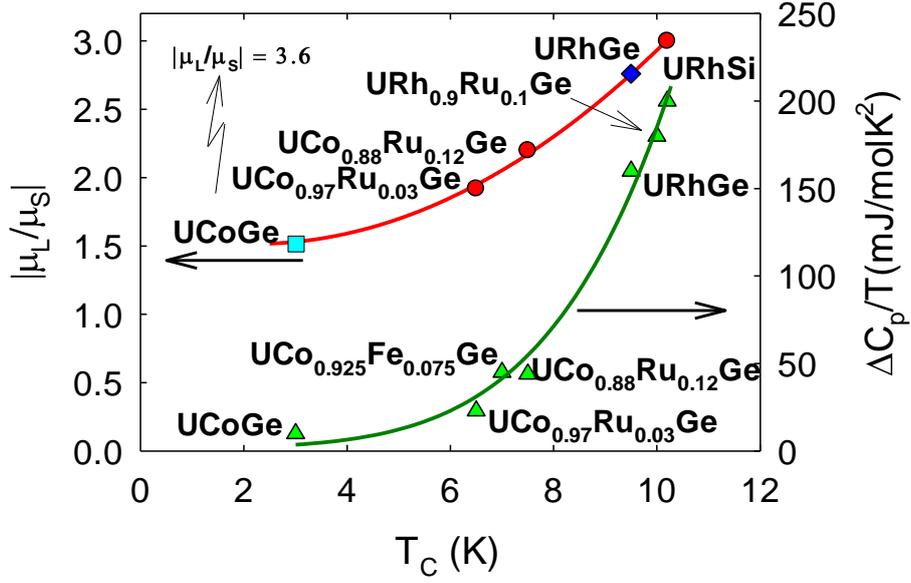

FIG. 6. (Color online) Evolution of the $|\mu_L^U/\mu_S^U|$ ratio and of the magnitude of heat capacity jumps $\Delta C_p/T$ with $T_C$ in the TiNiSi-type ferromagnetic UTX compounds. The blue point suggests the expected value of the $|\mu_L^U/\mu_S^U|$ ratio in URhGe where experimental data are still missing. The cyan point suggests the expected value of the $|\mu_L^U/\mu_S^U|$ ratio for the UCoGe respecting $\Delta C_p/T$ scaling. The URhSi value of $|\mu_L^U/\mu_S^U|$ was taken from data available in Refs.[25], $\Delta C_p/T$ are taken from Refs.[3, 13, 42]. The solid lines are guides to the eye.

The observed values of $\mu_L^U$ and $\mu_S^U$ in Table III are far from the free ions values. The $|\mu_L^U/\mu_S^U|$ ratio carries indirect information about the hybridization strength within the selected systems[52]. Here, the smaller $|\mu_L/\mu_S|$ reflects the strong hybridization with the *s*, *p*, *d* valence states and the direct overlap of the 5*f* wave functions (intermediate coupling). Generally, a high value of $|\mu_L^U/\mu_S^U|$ reveals a higher localization of the 5*f* electrons: the $\mu_L^U$ density in U intermetallics is usually distributed closer to the ion center than the spin density[53, 54]. For example in ferromagnetic UAsSe[55] with localized 5*f* states the ratio of $|\mu_L^U/\mu_S^U| = 3.2$, but, even there absolute values of and $\mu_S^U$ and $\mu_L^U$ are far from the free ions values. We have plotted the $|\mu_L^U/\mu_S^U|$ ratios as a function of $T_C$ (see Fig. 6). There is a clear increase of the $|\mu_L^U/\mu_S^U|$ ratios from UCo$_{1-x}$Ru$_x$Ge toward URhSi. Thus the hybridization strength as delocalization mechanism is the lowest in ferromagnetic URhSi where the 5*f* states are much localized within the TiNiSi-type UTX family. Since the behavior of UCoGe is controversial, we have only indicated in Fig. 6 an expected value of $|\mu_L^U/\mu_S^U|$, following the scaling with $\Delta C_p/T$ described below. The $|\mu_L^U/\mu_S^U|$ ratios for Ru substituted UCoGe single crystals lie between the expected UCoGe and URhSi ones, signaling that the 5*f* states are more localized than in the parent compound UCoGe and that the hybridization strength as delocalization mechanism is weaker. However, the low absolute values of $\mu_L^U$ and $\mu_S^U$ in comparison to UAsSe indicate that UCo$_{0.97}$Ru$_{0.03}$Ge and UCo$_{0.88}$Ru$_{0.12}$Ge and even URhSi can be classified as itinerant ferromagnets[56].

Our finding of a gradual localization of the 5*f* states in UCo$_{0.97}$Ru$_{0.03}$Ge and UCo$_{0.88}$Ru$_{0.12}$Ge deduced from $|\mu_L^U/\mu_S^U|$ ratios is in very good agreement with the relevant physical quantities obtained from our macroscopic measurements: We have found systematic scaling of the heat capacity jumps $\Delta C_p/T$ at $T_C$ for TiNiSi-type UTX compounds – see Fig.6. The highest $\Delta C_p/T$ has been found as expected for URhGe and URhSi. The proportional magnetic entropy, having a similar trend like $\Delta C_p/T$, is very weak even in URhSi suggesting high $5f$ electron itineracy[56] in agreement with PND data. Consistent with this, the value of $S_{mag}$ of UCoGe (only 3% of RLn2)[3] is very close to zero expected for true weak itinerant ferromagnets[5]. The found scaling of $\Delta C_p/T$ as a function of $T_C$ is subject to theoretical calculation within the SCR theory[57, 58] and will be published as a separate work.

The gradual localization of the $5f$ states is also reflected in the shape of the magnetization loops of UCo$_{0.97}$Ru$_{0.03}$Ge and UCo$_{0.88}$Ru$_{0.12}$Ge which are significantly more rectangular than that of parent UCoGe (Fig. 1) and the value of $\mu_{spont.}$ is almost 3 times higher in UCo$_{0.88}$Ru$_{0.12}$Ge than in UCoGe. Nevertheless, the magnetic moment of UCo$_{0.88}$Ru$_{0.12}$Ge at 1.8 K and 7 T is still only $\approx$ 10 % of the magnetic moment of free $U^{3+}$ or $U^{4+}$ ion. We also find that the magnitude of the spin fluctuations is significantly reduced in substituted compounds compared to that of UCoGe due to $5f$ states localization. This is evident from the behavior of the magnetization along the $b$-axis, which is hard like the a-axis magnetization and also from the weaker knee in electrical resistivity data at 40 K along the c-axis compared to that in parent UCoGe[32, 43]. The observed suppression of the ferromagnetic fluctuations in the substituted samples with additional electron-magnon scattering processes is most likely the reason of the immediate disappearance of superconductivity in the substituted systems.

On the basis of our PND data in combination with the macroscopic measurements, we conclude that the values of $T_C$ and $\mu_{spont.}$ in TiNiSi-type UTX compounds are determined by the strength of the $5f$-$d$ hybridization and the corresponding localization of the $5f$ states. Our PND results support the scenario that the initial stabilization of FM in URh$_{1-x}$Co$_x$Ge is realized via a growth of uranium orbital moment $\mu_L^U$ and the consequent localization of the $5f$ states on U site. This can also be generalized to other FM alloys with TiNiSi-structure such as UCo$_{1-x}$Fe$_x$Ge[13], URh$_{1-x}$Co$_x$Ge[59] and URh$_{1-x}$Ru$_x$Ge[17].

Our scenario is supported by the theoretical model proposed by Silva Neto et. al.[65] considering hybridization as a function of varying width and positions of the ligand $d$- and uranium $5f$-band to explain the origin of the observed FM dome in the URh$_{1-x}$Co$_x$Ge. A similar conclusion was also found for the URh$_{1-x}$Ru$_x$Ge system by XPS study where enlarged DOS at $E_F$ was found to be due to hybridized $4d$-$5f$ band[44]. The conclusions of the XPS study of URh$_{1-x}$Ru$_x$Ge also explain our observation of the enhanced value of $\gamma$ coefficient in UCo$_{0.97}$Ru$_{0.03}$Ge and UCo$_{0.88}$Ru$_{0.12}$Ge single crystals where non-Fermi liquid state develops in the boundary of ferromagnetism at x = 0.31%[14].

Finally we would like to discuss the relative values of the U and Co magnetic moments in our substituted compounds and compare them to those reported for UCoGe[21]. Generally, the existence of a weak induced parallel moment on the transition metal site is a common feature observed in all other so far studied UTX intermetallic compounds e.g. URhSi[25], UCoAl[26, 27], URhAl[28] or URuAl[29]. In contrast, a previous PND study[21] performed on the parent UCoGe suggests the existence of a very large Co magnetic moment antiparallel to the uranium moment in high magnetic field. This result is in disagreement with the recent XMCD results reported on UCoGe where parallel U and Co moments where found[30, 60]. The $|\mu_L^U/\mu_S^U|$ ratio from the XMCD spectra is almost field independent and has magnitude similar to that of our substituted compounds[30]. The $\mu_S^{Co}$ of the substituted compounds also scales well with the data obtained from the XMCD work[30].

Considering the fact that the magnetic moment at the cobalt site $\mu_S^{Co}$ is predominantly induced by the spin part of uranium moment[61] $\mu_S^{Co}$ can be used as a local probe for spin moment at the uranium site. As shown in Table III, the $\mu_S^{Co}/\mu_S^U$ ratios are similar within the experimental accuracy for both substituted compounds, justifying our assumption.

## V. Conclusions

We have successfully carried out PND experiments in two Ru substituted UCoGe single crystals to explore the microscopic mechanism of anomalous growth of $T_C$ and $\mu_{spont.}$ and the observed FM dome in UCo$_{1-x}$Ru$_x$Ge. We have shown that Ru substitution (x $\leq$ 0.1)[14] seems to lead to an increase of the U orbital moment $\mu_L^U$ (and consequently the total U moment), which reflects a certain localization of the 5$f$ electrons as manifested by a corresponding growth of the $|\mu_L^U/\mu_S^U|$ ratio. We may conclude that the appearance of the FM dome in UCo$_{1-x}$Ru$_x$Ge is due to a change of the degree of 5$f$-ligand hybridization which determined the degree of localization of the U 5$f$ states.

Our PND study confirmed that the initial grow of $T_C$ and $\mu_{spont.}$ (x $\leq$ 0.12) in UCo$_{1-x}$Ru$_x$Ge is due to weakness of hybridization strength as delocalization mechanism which supports the scenario suggested in Ref.[14]. A surprising scaling of the $|\mu_L^U/\mu_S^U|$ and heat capacity jumps $\Delta C_p/T$ at $T_C$ (and magnetic entropy $S_{mag}$) (Fig. 6) allows a generalization of gradual hybridization strength for all ferromagnetic UTX compounds adopting the orthorhombic TiNiSi–type crystal structure which clearly corroborates the scenario for UCoGe that it is a very weak itinerant ferromagnet with strongly delocalized 5$f$ electrons.

The study has further shown that the most localized 5$f$ electrons within the series are in URhSi and URhGe, even so the localization of the 5$f$ states is still far from typical local moment ferromagnets like UAsSe. The gradual localization of the 5$f$ states in the substituted compounds is also in very good agreement with the relevant physical quantities obtained from our macroscopic measurements which reflect the gradual reduction of spin fluctuations with increasing x up to 0.12. Magnetization loops are more rectangular with higher $\mu_{spont.}$, lower high-field susceptibility, higher jumps $C_P$ jumps at $T_C$. In addition, the associated magnetic entropy $S_{mag}$ in specific heat and also knee at $\approx$ 40 K in the electrical resistivity can be attributed to a reduction of spin fluctuations with increasing Ru content. The suppressed ferromagnetic fluctuations are probably (besides the substitutional disorder) responsible for the sudden loss of superconductivity in the substituted systems.

The parallel alignment of the U and Co moments observed in UCo$_{0.97}$Ru$_{0.03}$Ge and UCo$_{0.88}$Ru$_{0.12}$Ge are in agreement with the results of the Compton scattering and XMCD experiments on UCoGe at temperatures of the normal ferromagnetic state. Further detailed PND and XMCD studies are necessary to reveal the origin of high field and low temperature magnetic state of UCoGe[21].

**Acknowledgements**

Authors would like to thank S. Kambe, K. Kaneko, N. Tateiwa and Z. Fisk for fruitful discussion of the results. This work was supported by the Czech Science Foundation no. P204/12/P418 and by the Charles University in Prague, project GA UK No.720214. Experiments performed in MLTL (see: http://mltl.eu/) were supported within the program of Czech Research Infrastructures (project LM2011025). The neutron experiment in ILL is a part of research project LG14037 financed by the Ministry of Education, Youth and Sports, Czech Republic. The XMCD experiment was performed under the proposal No. 2014A3821 and 2014B3821 of SPring-8 BL23SU and was financially supported by JPSP KAKENHI Number 25800207.